\documentclass[fleqn,usenatbib]{mnras}

\usepackage{newtxtext,newtxmath}

\usepackage[T1]{fontenc}

\DeclareRobustCommand{\VAN}[3]{#2}
\let\VANthebibliography\thebibliography
\def\thebibliography{\DeclareRobustCommand{\VAN}[3]{##3}\VANthebibliography}

\usepackage{graphicx}	
\usepackage{amsmath}	

\title[SFR Enhancements in the Post-merger phase]{Galaxy Evolution in The Post-Merger Regime I -- Most merger-induced in-situ stellar mass growth happens post-coalescence}

\author[Ferreira et al.]{Leonardo Ferreira$^{1}$,\thanks{E-mail: lferreira@uvic.ca} 
Sara L. Ellison$^{1}$, 
David R. Patton$^{2}$, 
Shoshannah Byrne-Mamahit$^{1}$, 
Scott Wilkinson$^{1}$,
\newauthor
Robert Bickley$^{1}$, 
Christopher J. Conselice$^{3}$, 
Connor Bottrell$^{4}$  \\ 
$^{1}$ School of Physics and Astronomy, University of Victoria, Victoria, BC, Canada\\
$^{2}$Department of Physics and Astronomy, Trent University, 1600 West Bank Drive, Peterborough, ON K9L 0G2, Canada \\
$^{3}$Jodrell Bank Centre for Astrophysics, University of Manchester, Oxford Road, Manchester UK \\
$^{4}$International Centre for Radio Astronomy Research, University of Western Australia, 35 Stirling Hwy, Crawley, WA 6009, Australia \\
}

\date{Accepted XXX. Received YYY; in original form ZZZ}

\pubyear{\the\year{}}

\begin{document}
\label{firstpage}
\pagerange{\pageref{firstpage}--\pageref{lastpage}}
\maketitle

\begin{abstract}
Galaxy mergers can enhance star formation rates throughout the merger sequence, with this effect peaking around the time of coalescence. However, owing to a lack of information about their time of coalescence, post-mergers could only previously be studied as a single, time-averaged population. We use timescale predictions of post-coalescence galaxies in the UNIONS survey, based on the Multi-Model Merger Identifier deep learning framework (\textsc{Mummi}) that predicts the time elapsed since the last merging event. For the first time, we capture a complete timeline of star formation enhancements due to galaxy mergers by combining these post-merger predictions with data from pre-coalescence galaxy pairs in SDSS. Using a sample of $564$ galaxies with $M_* \geq 10^{10} M_\odot$ at $0.005 < z < 0.3$ we demonstrate that: 1) galaxy mergers enhance star formation by, on average, up to a factor of two; 2) this enhancement peaks within 500 Myr of coalescence; 3) enhancements continue for up to 1~Gyr after coalescence; and 4) merger-induced star formation significantly contributes to galaxy mass assembly, with galaxies increasing their final stellar masses by, $10\%$ to $20\%$ per merging event, producing on average $\log(M_*/M_\odot) = {9.56_{-0.19}^{+0.13}}$ more mass than non-interacting star-forming galaxies solely due to the excess star formation. 

\end{abstract}




\section{Introduction}

Galaxies primarily assemble their stellar mass by either forming new stars from available gas \citep{Madau2014} or combining their stellar content with neighbouring companions through merging \citep{Duncan2019}. A discussion on the relative importance of these effects to the build up of galaxies throughout cosmic history is still open \citep{Rodriguez-Gomez2016}, mainly due to the lack of tight constraints on galaxy merging histories \citep{Casteels2014, Duan2024}.

Usually addressed as separated entities, mass assembly through star-formation (in-situ) and merging (ex-situ) are mechanisms that are correlated to some degree. Galaxy mergers that are star-forming exhibit enhanced star-formation rates when compared to their non-interacting counterparts \citep{scudder2012, patton2013, Scott2014, Violino2018, Bickley2021, Garay-Solis2023}. Hence, interactions play an important role in fostering the conditions for increased star-formation, either by increasing the inflow of pristine gas from the circumgalactic medium \citep{Hani2016} or by re-arranging the already-present gas content \citep{NCA2004, thorp2019}.

As galaxy interactions unfold over billion year long timescales \citep{Rodriguez-Gomes2015}, in two distinct phases (pre-coalescence and post-coalescence), the impact on merging to star-formation is not instantaneous and evolves along this sequence \citep{Ellison2013, Bickley2022}. The importance of the pre-coalescence phase has been thoroughly studied \citep{Ellison2008, scudder2012, patton2013, Patton2016} by using the physical separation and relative velocities between companions, together with its dynamical implications, as a proxy for time until coalescence \citep{Snyder2015, Patton2024}. It has been shown that star-formation rates in nearby mergers start to be enhanced as soon as paired galaxies are as close as $150$~kpc, with this effect increasing over time until the closest separations we can possibly resolve spatially \citep{patton2013}. 

However, this dynamical assessment at the post-coalescence phase (post-mergers) is not as straightforward or even possible. Morphological disturbances present in post-merger galaxies are the main evidence used for their identification \citep{NairandAbraham2010, Ellison2013, WPR2020}, and although these signatures have timescales associated with them (usually based on how strong/clear they are), observational and cosmological effects shroud their temporal relationship \citep{Wilkinson2024}. Thus, the effect on the post-merger phase to star formation in observations has been only explored averaged for the whole population. And yet, there is evidence that these enhancements along the merger sequence peak at the post-merger phase \citep{Ellison2013, Bickley2022}. Ultimately, the real impact of merging to star formation and galaxy mass assembly can not be summarized if the post-merger phase is unconstrained temporally, as these enhancements can not be translated to how much mass this adds to each merging galaxy. 

Recently, \citet{Koppula2021} and \citet{Pearson2024} showed that it is possible to separate post-mergers temporally by their morphology when combining cosmological simulations and deep learning. In \citet{Ferreira2024a}, we developed the Multi Model Merger Identifier (\textsc{Mummi}), and showed that making these models robust with respect to the overall purity on billion year long timescales allow us to apply them to observations reliably. Ferreira et al. (in prep) introduces a framework to infer post-coalescence  timescales based only on imaging, in four distinct time bins spanning a $1.76$ Gyr time window after the merging event. These models were trained and validated with the IllustrisTNG cosmological simulations \citep{Nelson2018, Pillepich2018}.

The work presented here introduces a series of studies exploring the temporal evolution of post-mergers as presented in the catalogs of \citet[][and in prep.]{Ferreira2024a} in the Ultraviolet Near Infrared Optical Northern Survey (UNIONS). Here we assess the temporal evolution of star formation enhancements throughout the whole merger sequence, from the pair phase until billion years after coalescence. We report that not only do these enhancements peak immediately after coalescence, but they only go back to normal after 1 Gyr. We investigate how much mass is added to galaxies due to star formation induced from merging.



We assume the same cosmological model used by IllustrisTNG, which is consistent with the \citet{Planck2018} results that show $\Omega_{\Lambda,0} = 0.6911$, $\Omega_{m,0} = 0.3089$, and $h = 0.6774$.





\section{DATA}\label{sec:data}

\begin{figure*}
    
    \centering
    \includegraphics[width=1\textwidth]{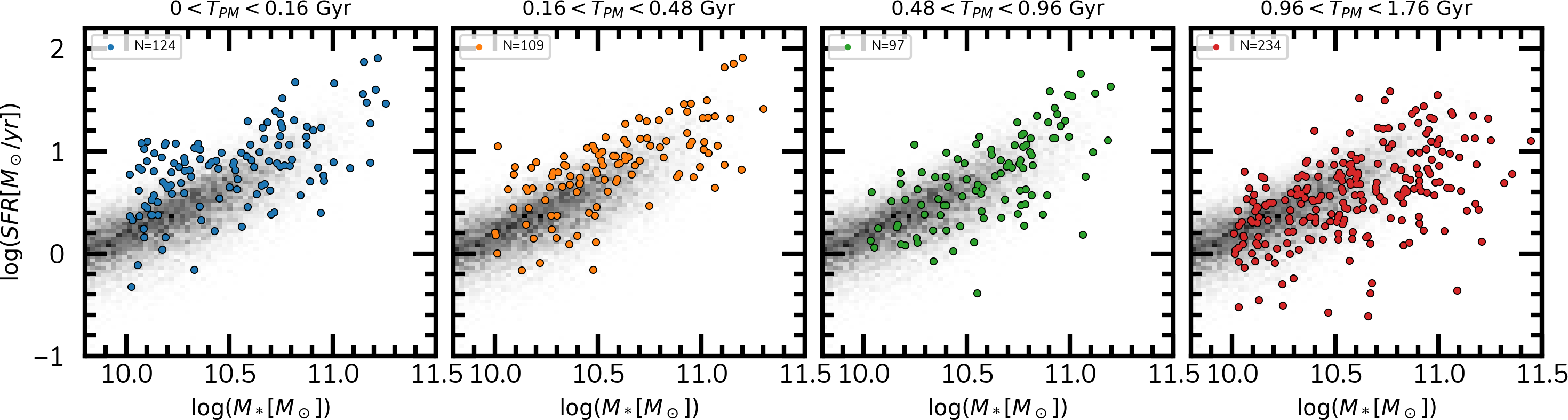}
    \caption{\textbf{SFR vs. stellar mass of the post-merger sample.} Each panel shows the distribution of SFRs and stellar masses for the corresponding time bin in our post-merger sample in coloured points, with $0 < T_{PM} < 0.16$ Gyr in blue, $0.16 < T_{PM} < 0.48$ in orange, $0.48 < T_{PM} < 0.96$ in green, and $0.96 < T_{PM} < 1.76$ in red, respectively. We plot in the background as a gray 2D histogram the complete control pool of \textsc{Mummi} non-mergers to be matched to the post-merger sample, representing the star forming main sequence. It is clear that the shortest timescales exhibit the highest positive offsets.}
    \label{fig:sample}
\end{figure*}

In this Letter we explore the temporal evolution of galaxy merger induced star formation enhancements in UNIONS galaxies on the star-forming main sequence. We use the r-band of the UNIONS consortium of wide-field imaging surveys in the Northern hemisphere. The r-band imaging was taken at the 3.6-metre Canada-France–Hawaii Telescope (CFHT) on Maunakea, mapping $4861$ deg$^2$ of the northern sky, reaching a 5$\sigma$ depth in $28.4$ mag arcsec$^{-2}$. These data provides the best wide-survey imaging available to-date for merger identification as many low-surface brightness features not detectable in Sloan Digital Sky Survey (SDSS) images \citep{York2000} are clearly detected with it.

To build this timeline, from the pre-coalescence close pair regime all the way to the post-coalescence phase, we require merger identification for each stage, as well as a sample of non-interacting control galaxies. Additionally, as our goal is to track the evolution of the star formation rates due to merging, and control the effects of redshift and stellar mass evolution, we use redshifts from the SDSS Data Release 7 (DR7) and stellar masses and star formation rates from the MPA-JHU catalog \citep{Kauffmann2003}.

\subsection{Pairs Sample}

To track the effects of galaxies at the onset of an interaction, we require a sample of spectroscopically confirmed galaxy pairs that have relatively close projected separations and with small relative velocities along the line of sight, indicating they are likely to coalesce in the future. To do this, we rely on the catalog constructed by \citet{Patton2016}. In this sample, the closest companion within $2$~Mpc of each galaxy in the SDSS DR7 was identified based on stellar mass ratios within a factor of 10 and a velocity separation of $\Delta v < 1000$~km~s$^{-1}$. For this study, as we are not concerned with the long-range effects of the pair phase and given that evidence shows SFR enhancements are minimal at separations greater than $100$~kpc \citep{patton2013}, we have refined the sample with a stricter cut. We selected only pairs of galaxies where the nearest companion lies within $100$~kpc and has a velocity separation of $\Delta v < 300$~km~s$^{-1}$. Additionally, to ensure reliable post-merger identification (Ferreira et al. in prep), we restrict our selection to massive galaxies with stellar masses of $\log(M_*/M_\odot) \geq 10$, and apply an upper redshift cut of $z < 0.3$ (see next sub-section).

Since we are focusing exclusively on star forming galaxies, we use BPT diagram classifications from \citet{Kauffmann2003}, ensuring a signal-to-noise ratio of at least 3 in each emission line. After applying all these constraints, our resulting sample has 2805 pair galaxies.

\subsection{Post-merger Sample}

Post-coalescence galaxies require different identification methods compared to pairs, as they must be classified based on morphological disturbances caused by the merging event. Historically, this classification has been performed visually or through quantitative automated methods \citep{Conselice2003b, NairandAbraham2010, Kartaltepe2015}. However, these methods often struggle with a poor balance between purity and completeness \citep{Wilkinson2024, Bickley2024b}, and they are unable to accurately infer timescales apart from some isolated cases \citep{Privon2013}.

More recently, there have been successful attempts to build physically motivated frameworks that predict merging timescales from morphology alone, using information from cosmological simulations \citep{Koppula2021}. However, models have been shown to work in theory, but there has yet to be an application to quantitatively characterize mergers in the post-coalescence regime  \citep{Pearson2024}. In the next section we briefly describe the Multi Model Merger Identifier (\textsc{Mummi}), and our approach to forward model cosmological simulations to the observational domain to classify mergers and to measure their timescales reliably.

\subsubsection{Multi Model Merger Identifier \textsc{Mummi}}

Here we rely on classifications by \textsc{Mummi} \citep[][Ferreira et al. in prep]{Ferreira2024a}, a large ensemble of deep learning models tailored to work with UNIONS data. \textsc{Mummi} works hierarchically in three subsequent tasks: merger identifications (STEP 1), stage classification (pair/post-merger, STEP 2) and timescale predictions (STEP 3). We briefly review the process here but refer the reader to \citet{Ferreira2024a} and Ferreira et al. (in prep) for more details.

\begin{itemize}
    \item STEP 1: A collection of 20 different deep learning models (CNNs and Vision Transformers) is used to classify galaxies as mergers or non-mergers. Each of the 20 models casts a vote, and we use the number of votes to indicate \textsc{Mummi}'s certainty in its classification. Usually, a simple majority of 11 out of 20 models is used to infer positive merger status, but other configurations are possible.

    \item STEP 2: a pair of models decides whether the mergers classified in STEP 1 are pairs or post-mergers. \textsc{Mummi} achieves $96\%$ success rate at this task. In the present work, we do not use the pair classifications for our pair sample but use them only to remove contamination introduced by very close pairs from the post-merger sample.

    \item STEP 3: The third step (described in detail in Ferreira et al. in prep) uses a new ensemble of 20 deep learning models to predict a timescale for the post-mergers classified in STEP 2. Instead of a regression task, this step is solved through a classification procedure. In Ferreira et al. (in prep) we show that we can reliably classify post-mergers into 4 different timescale bins: $0 < T_{PM} < 0.16$~Gyr, $0.16 < T_{PM} < 0.48$~Gyr, $0.48 < T_{PM} < 0.96$~Gyr, and $0.96 < T_{PM} < 1.76$~Gyr. Although these time bins are sparsely sampled, they are enough for us to track the temporal evolution of physical processes in the post-merger phase for the first time. \textsc{Mummi} achieves $\sim 75\%$ precision at this task.
\end{itemize}


We use the publicly available \textsc{Mummi} catalogs covering the overlap between UNIONS DR5 and SDSS DR7 \citep[][Ferreira et al. in prep]{Ferreira2024a}, which include classification information for all three STEPS in the pipeline. From $235,354$ available galaxies, we select those that are massive with $\log(M_*/M_\odot) \geq 10$, are in the redshift range of $0.005 < z < 0.3$, and are classified as SF by \citet{Kauffmann2003}, with $S/N > 3$. From these, we select as post-mergers all remaining galaxies with at least 11 votes in STEP 1\footnote{We also explored different post-merger selections with higher vote thresholds. We do not find any significant trend with higher votes, apart from smaller overall post-merger samples.}, post-merger probabilities of $p(x) > 0.5$ in STEP 2, and probability distributions in STEP 3 with a clear peak. This results in 564 post-merger galaxies with time predictions, distributed across the time bins as 124, 109, 97, and 234, respectively.

\subsection{Control Samples}

To investigate the impact of merging on star formation, we create control samples for each merger stage. We use the galaxies from the initial selection, but that were not given merger classifications.

For pairs, we select all galaxies that do not have close companions within $r_p < 100~$kpc, with no restriction on relative velocities. Additionally, we apply the same mass and BPT cuts as used for the pair galaxies, resulting in a pool of 115,907 possible controls.

For the post-mergers, our control pool includes all cases with 2 or fewer votes in STEP 1 as our non-merging sample, applying the same constraints on the BPT diagram, mass, and redshift.  This yields a sample of 26,354 non-interacting galaxies.

Finally, we find matching control galaxies to each merger galaxy based on stellar mass and redshift, using two highly restrictive iterative Kolmogorov–Smirnov test \citep{KSTEST}, halting the matching once a p-value of under $99\%$ is found in either stellar mass or redshift. For the pairs, we match 9 controls per galaxy, whereas for the post-mergers, the number of controls varies per bin, as matching is done independently for each time interval. We have 16, 43, 29, and 28 controls per galaxy, respectively, from shortest to longest.

\section{RESULTS}

Here, we create a complete observational timeline of star formation enhancements in star forming galaxies from the start of an interaction, through the coalescence of the stellar masses. For a view of the impact of timescales into star formation quenching see \citet{ellison:submitted}.

In Figure~\ref{fig:sample},  we show the distributions of stellar masses and SFRs for our post-merger sample in each of the four time bins $0 < T_{PM} < 0.16$~Gyr, $0.16 < T_{PM} < 0.48$~Gyr, $0.48 < T_{PM} < 0.96$~Gyr, and $0.96 < T_{PM} < 1.76$~Gyr, with the blue, orange, green, and red points, respectively. For comparison, we also display the entire distribution of the control pool, which was matched to the post-mergers, in the shaded 2D histograms. We display only the post-merger sample because the evolution of SFRs in the pair phase has been extensively studied in the literature. Comparing the positions of the mergers with respect to the controls in Figure~\ref{fig:sample} reveals that post-mergers with shorter $T_{PM}$ have on average higher offsets from the star forming main sequence. Each subsequent timescale shows smaller positive offsets and greater negative offsets. 

\begin{figure}
    \centering
    \includegraphics[width=0.45\textwidth]{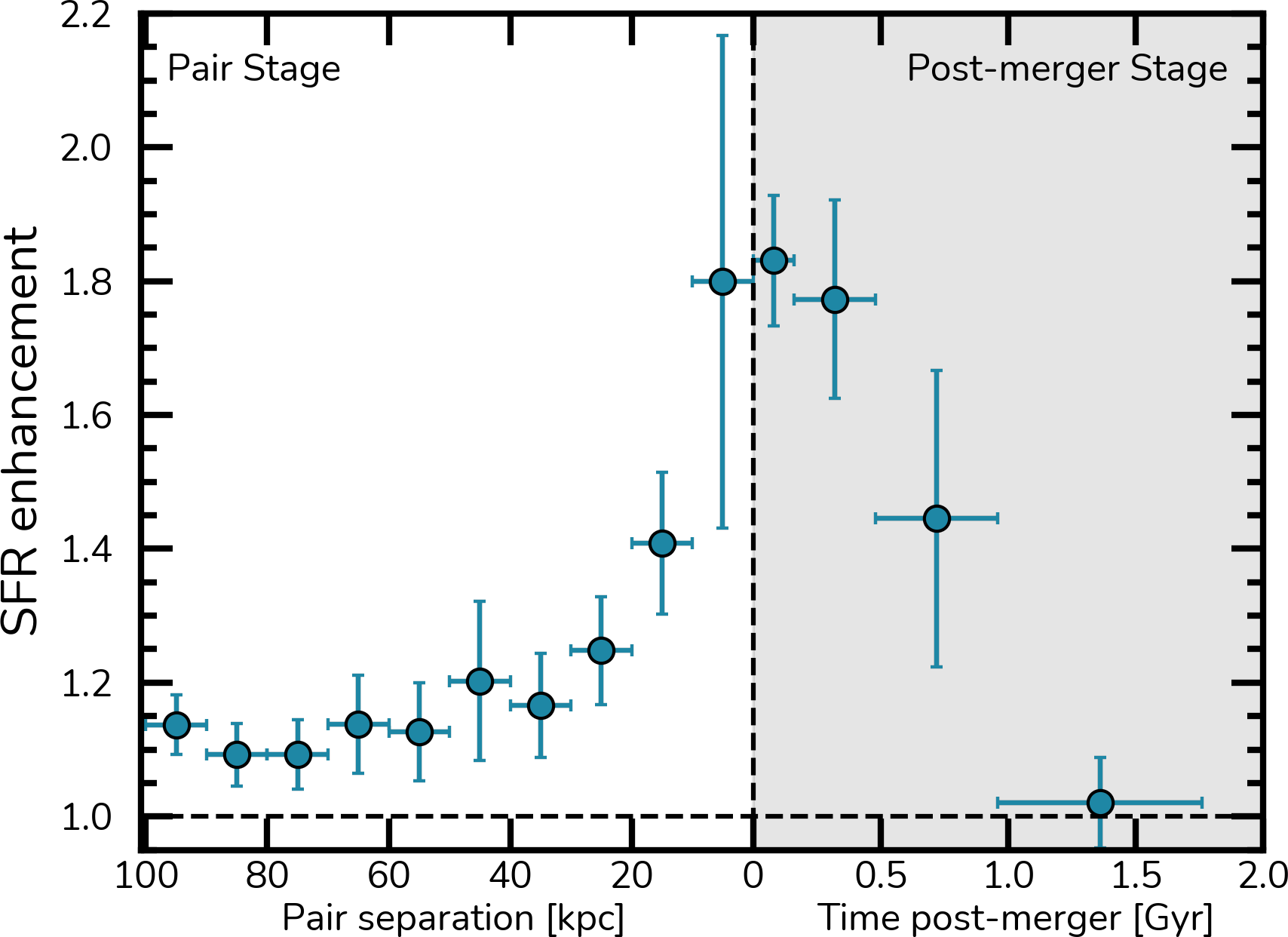}
    \caption{\textbf{SFR Enhancements along the merger sequence.} We show the evolution of SFR enhancements in the pair stage (left) and post-merger stage (right). For the pairs, we use their proximity as a proxy for a time until merger, while for post-mergers we use \textsc{Mummi} timescale predictions. Galaxies display increasing enhanced SFRs with decreasing proximity with their companions, with this effect peaking at around the time of coalesence. SFRs can be enhanced as much as 2x when compared to selected controls. In the post-merger phase, these enhancements decay rapidly over a 1 Gyr time window, going back to normal at $0.96 < T_{PM} < 1.76$. }
    \label{fig:enhancementtimeline}
\end{figure}

Using the samples of pre-coalescence and post-coalescence galaxy mergers selected in \S~\ref{sec:data}, along with their matched controls in redshift and stellar mass, we have created the first observational timeline of merger-induced star formation rate  enhancements across the galaxy merger sequence. In Figure \ref{fig:enhancementtimeline}, we show the evolution of SFR enhancements in two regimes. This timeline can be read from left to right, showing galaxies in the pair phase gradually decaying in their orbits until coalescence, followed by dynamic settling into a single galaxy in the post-merger phase.

First, in the pair phase (left side of Fig. \ref{fig:enhancementtimeline}), we track galaxies from a minimum pair separation of $r_p \leq 100$ kpc to the time just before coalescence at separations of $10 > r_p > 0$ kpc. At $r_p \sim 100$ kpc, small enhancements are already present,  gradually increasing with decreasing pair separation, reaching up to an enhancement of $1.8{\pm0.4}$ just before coalescence, in precise agreement with \citet{Bickley2021}. The last data point in the pair sequence is affected by fiber collision issues \citep{Patton2016}, resulting in a smaller sample and hence bigger uncertainties.

Previous work has shown that the small but present enhancements at $\sim 100$~kpc are not detected beyond $\sim 150$ kpc \citep{patton2013}. To track star formation enhancements up to these longer separations would require a more involved and complex control matching scheme that takes into account environmental effects, as these play an important role at $ > 100$ kpc \citep{patton2013}. As we will show in \S~\ref{subsec:growth}, these small enhancements are negligible to the merger-induced mass growth explored here. Additionally, we note that pair separation can be used as a proxy for time until merger. For example, using a sample of galaxy pairs from the IllustrisTNG cosmological simulations, \citet{Patton2024} find that the median 3D pair separation decreases by 55 kpc/Gyr during the 5 Gyr period leading up to the merger. Using the same pairs sample, the median projected separation is estimated to decrease by 42 kpc/Gyr (D. Patton private communication). Galaxy mergers occur through various orbital configurations, and this conversion factor should not be applied on a galaxy-by-galaxy basis. However, it describes a statistical average of orbital decay over time for a large sample of galaxies.  

Second, in the post-merger phase (right side of Fig. \ref{fig:enhancementtimeline}), as determined from \textsc{Mummi} classifications and post-merger time predictions, the data are divided into one of four time bins. The greatest enhancements are found immediately after coalescence, with enhancements of $1.9\pm0.1$ similar to values found at the end of the pair phase. These enhancement levels continue for up to $\sim 0.5$ Gyr. Beyond $\sim 0.5$ Gyr we find that star formation enhancements decay rapidly from $0.48 < T_{PM} < 0.96$~Gyr, with a value of $1.5\pm0.25$ to $1\pm0.05$ at $T_{PM} > 0.96$ Gyr, returning to the same baseline as the controls. Our results show longer enhancements when compared to results from IllustrisTNG, lasting an extra $0.5$~Gyr \citep{Hani2020}.



\subsection{Merger-induced Net Stellar Mass}\label{subsec:growth}

The primary advantage of revealing the complete temporal evolution of star formation enhancements in star-forming galaxies is that it allows us to estimate how much new in-situ stellar mass is induced by the merger event. Here, we explore the amount of mass generated from the excess star formation rates in mergers compared to control galaxies.


Since each merger in our sample (both pre- and post-coalescence) is matched to a number of control galaxies, we can estimate the median SFR of the controls, and a $\Delta SFR$ for each merger galaxy. Using this, we can compute the extra mass in a given galaxy, $\Delta M{_*}$ in units of $M_\odot$, formed by integrating $\Delta SFR$ within their respective time bins as
\begin{equation}
    \Delta M_{*} = \sum_{n}^{i} \Delta SFR~\Delta{t_i}\quad[M_\odot]
\end{equation}{}
where $\Delta{t_i}$ is each time bin size. Then, by averaging the contribution of $\Delta M_{*}$ from all galaxies in each bin, we can estimate how much mass, $M_{*, \text{excess}}$, that period of star-formation enhancement adds, on average, per  galaxy.

\begin{figure}
    \centering
    \includegraphics[width=0.45\textwidth]{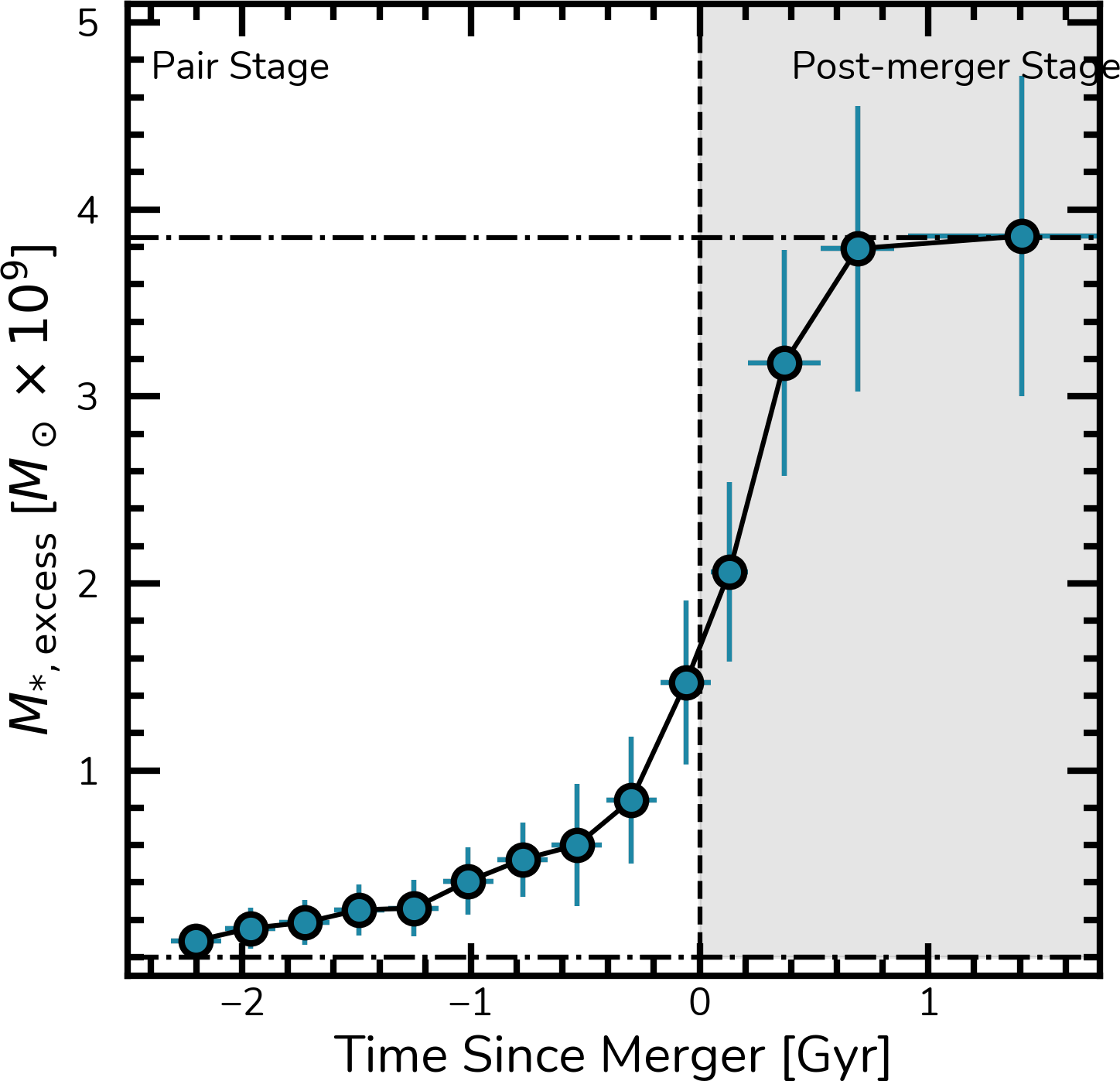}
    \caption{\textbf{Cumulative Mass Formed along the merger sequence.} We show how much cumulative stellar mass (current bin plus previous ones) is formed due to the enhanced star-formation along the complete merger sequence for all our sample. Merger-induced star-formation can form around $10^9 M_\odot$ by the end of the pair phase, and $10^{9.56} M_\odot$ in total by the end of the post-merger phase. This means that most mass formed due to enhancements from mergers are formed during the post-merger phase.}
    \label{fig:cum_total}
\end{figure}

In Figure~\ref{fig:cum_total} we show the cumulative effect of $M_{*, \text{excess}}$ in each time bin to the total mass formed across the merger sequence for our whole sample of $\log(M_*/M_\odot) > 10$ galaxies. As can be seen, the pair phase adds around $\sim1.5\times10^{9} M_\odot$, with around one third of that mass formed immediately before the coalescence. On the other hand, the merger-induced mass growth of the post-merger phase is of the order of $\sim2.5\times10^{9} M_\odot$ in around $\sim 500 $ Myr, almost double that of the pair-phase. In total, for the whole merger sequence, massive galaxies go through, on average, a merger-induced mass growth (only in-situ) of $\log(M_*/M_\odot) = {9.56_{-0.19}^{+0.13}}$. This is in very good agreement with the work done by \citet{Bottrell2024} within the IllustrisTNG simulations. Hence, most stellar mass formed from merger-induced star formation takes place in the post-merger phase which, until now, was an effect that could not be estimated observationally. 

\begin{figure}
    \centering
    \includegraphics[width=0.45\textwidth]{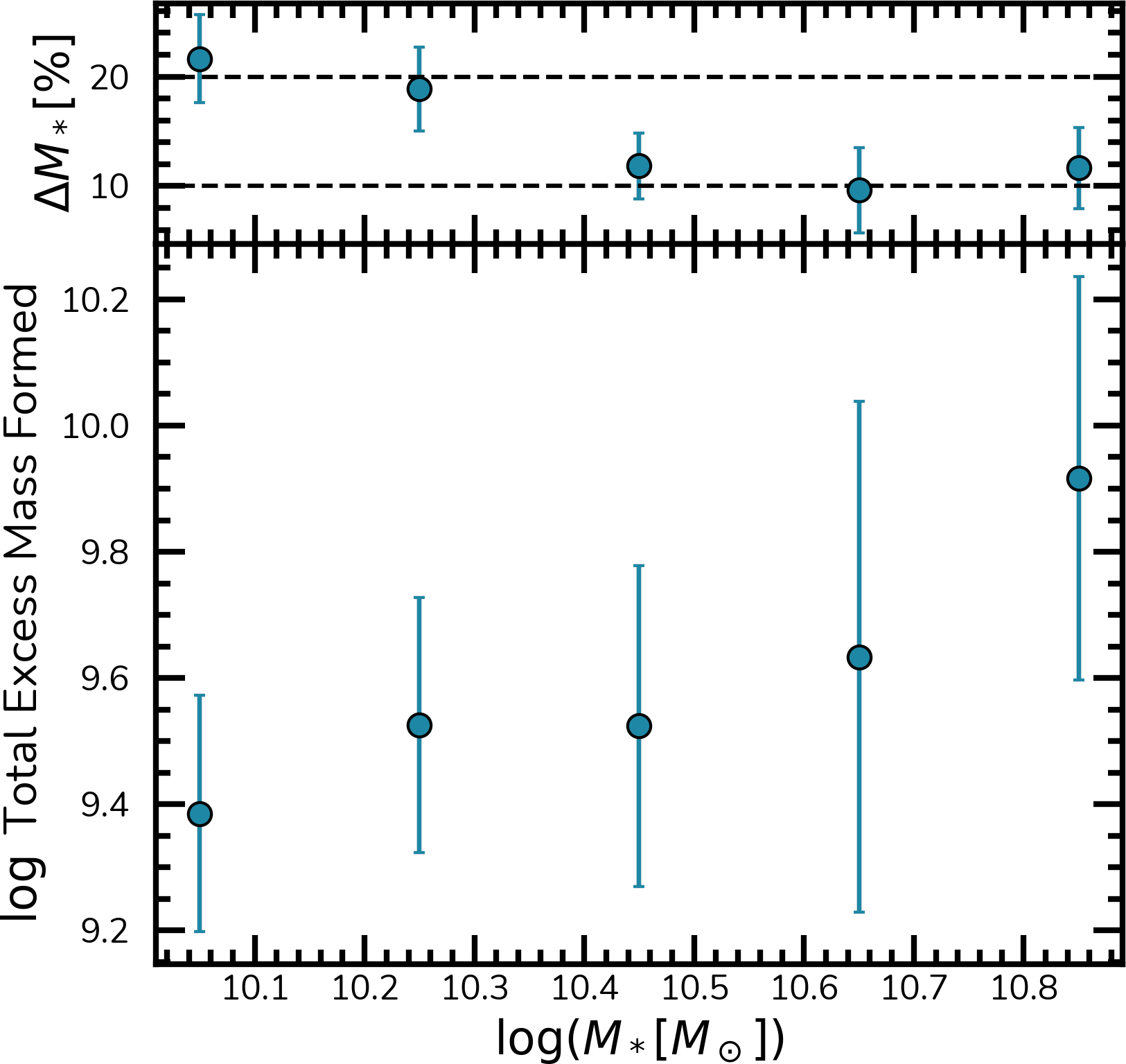}
    \caption{\textbf{Total cumulative mass formed in stellar mass bins.} Here we report the final amount of stellar mass formed from the excess star formation after the end of the merger sequence in small stellar mass bins $10.0 < \log(M_*/M_\odot) < 11.0$ (bottom), as well as the relative percentage mass increase for that given bin in the top frame. Galaxies form more stars due to merger-induced star formation based on their stellar masses, with $10^{11} M_\odot$ galaxies forming almost a order of magnitude more mass than $10^{10} M_\odot$ galaxies. We show that this effect increases a single galaxy's mass in the $20\%-10\%$ range.}
    \label{fig:cum_mass_bins}
\end{figure}

With Figure~\ref{fig:cum_total}, we can only indirectly infer how much new in-situ mass is added by the end of the merger relative to a specific galaxy mass as we do not have statistics to separate the sample in smaller mass bin across each timescale bin. To address this, in Figure~\ref{fig:cum_mass_bins} we report the total amount of mass formed in five stellar mass bins of $0.2$ dex in the range $10 < \log(M_*/M_\odot) < 11$ (bottom panel) and the fractional mass increase generated by this new mass (top panel). The amount of mass formed is significant across the whole mass range, with increases ranging from of $10\%$ to $20\%$. Its importance initially decreases from $20\%$ at around  $\log(M_*/M_\odot) \simeq 10.1$ to $\sim10\%$ at around $\log(M_*/M_\odot) \simeq 10.5$, and then subsequently stays at a constant level up $\log(M_*/M_\odot) \simeq 11$. This means that, to maintain a mass growth contribution of at least $10\%$ at this wide mass range, more massive galaxies tend to form more stars due to star formation enhancements, increasing from $10^{9.5}~M_*$ at $\log(M_*/M_\odot) \simeq 10.5$ to $10^10~M_*$ at $\log(M_*/M_\odot) \simeq 11$. This result is in general agreement with the stellar mass burst fractions measured by \citet{Reeves2024} at this same stellar mass range. We attribute the cause of this effect either due to stronger bursts at high mass, or by these more massive systems being preferentially higher mass ratio mergers, with more gas content.

\section{SUMMARY}

In this Letter we showcase for the first time a complete timeline of merger-induced star formation enhancements in galaxies along the star formation main sequence. This is done by combining a sample of spectroscopically confirmed paired galaxies in SDSS with post-merger timescale measurements from the Multi Model Merger Identifier (\textsc{Mummi}) of galaxies in UNIONS \citep[][in prep]{Ferreira2024a}. We combined these with a control sample of galaxies matched in redshift and stellar mass.

We capture the impact of merger-induced star formation enhancements from their onset, when galaxies are approaching each other with enhancements increasing as pair separation decreases, to the post-merger regime, where these enhancements decay over time until returning to nominal levels approximately one billion years after the merger. By separating post-merger classifications over time, we can track the evolution of enhancements from the closest separations in the pair phase to immediately after coalescence, observing how rapidly these enhancements evolve during the post-merger stage. In the past, the post-merger regime could only be investigated in an integrated manner without time constraints, despite the potential diversity within the selected populations.

Specifically, the $\sim2\times$ enhancements present at the end of the pair phase are also found for at least $500$ Myr after coalescence, then rapidly decaying over the next $500$ Myr, returning to no enhancements by 1 Gyr. This study captures the effect in its entirety for the first time, allowing us to understand not only how these enhancements evolve with time, but also to estimate the amount of mass growth solely due to merger-induced elevated star formation rates (in-situ). This star formation excess in mergers can increase stellar masses by, on average, $10\%$ in extremely massive galaxies and up to $20\%$ in less massive systems, making it nearly as significant as the stellar mass accreted from the companion galaxies..

While merging (ex-situ) and star formation (in-situ) are typically considered completely separate avenues for mass assembly, we show here that a significant fraction of in-situ star formation can actually be associated with merging. This raises the question of how much of the universe's stellar formation (i.e., the cosmic star formation history) is actually driven by merging. A detailed description of merging histories across cosmic time, combined with time-sensitive merger classifications, could help address this question.

\section*{Acknowledgements}

We are honored and grateful for the opportunity of observing the
Universe from Maunakea and Haleakala, which both have cultural,
historical and natural significance in Hawaii. This work is based on data obtained as part of the Canada-France Imaging Survey, a CFHT large program of the National Research Council of Canada and the French Centre National de la Recherche Scientifique. Based on observations obtained with MegaPrime/MegaCam, a joint project of CFHT and CEA Saclay, at the Canada-France-Hawaii Telescope (CFHT) which is operated by the National Research Council (NRC) of Canada, the Institut National des Science de l’Univers (INSU) of the Centre National de la Recherche Scientifique (CNRS) of France, and the University of Hawaii. This research used the facilities of the
Canadian Astronomy Data Centre operated by the National Research
Council of Canada with the support of the Canadian Space Agency.
This research is based in part on data collected at Subaru Telescope, which is operated by the National Astronomical Observatory of Japan.  We acknowledge the support of the Digital Research Alliance of Canada for providing the compute infrastructure to train the models presented here. SLE, DRP and LF gratefully acknowledge NSERC of
Canada for Discovery Grants which helped to fund this research. SJB and SW acknowledge graduate fellowships funding from the Natural Sciences and Engineering Research Council of Canada (NSERC); Cette recherche a été financée par le Conseil de recherches en sciences naturelles et en génie du Canada (CRSNG). CB gratefully acknowledges support from the Forrest Research Foundation.

\section*{Data Availability}

The catalogues used in this work were presented in \citet{Ferreira2024a} and Ferreira et al. (in prep), and are publicly available\footnote{\url{https://github.com/astroferreira/MUMMI_UNIONS}}.



\bibliographystyle{mnras}
\bibliography{paper} 

\begin{thebibliography}{}
\makeatletter
\relax
\def\mn@urlcharsother{\let\do\@makeother \do\$\do\&\do\#\do\^\do\_\do\%\do\~}
\def\mn@doi{\begingroup\mn@urlcharsother \@ifnextchar [ {\mn@doi@} {\mn@doi@[]}}
\def\mn@doi@[#1]#2{\def\@tempa{#1}\ifx\@tempa\@empty \href {http://dx.doi.org/#2} {doi:#2}\else \href {http://dx.doi.org/#2} {#1}\fi \endgroup}
\def\mn@eprint#1#2{\mn@eprint@#1:#2::\@nil}
\def\mn@eprint@arXiv#1{\href {http://arxiv.org/abs/#1} {{\tt arXiv:#1}}}
\def\mn@eprint@dblp#1{\href {http://dblp.uni-trier.de/rec/bibtex/#1.xml} {dblp:#1}}
\def\mn@eprint@#1:#2:#3:#4\@nil{\def\@tempa {#1}\def\@tempb {#2}\def\@tempc {#3}\ifx \@tempc \@empty \let \@tempc \@tempb \let \@tempb \@tempa \fi \ifx \@tempb \@empty \def\@tempb {arXiv}\fi \@ifundefined {mn@eprint@\@tempb}{\@tempb:\@tempc}{\expandafter \expandafter \csname mn@eprint@\@tempb\endcsname \expandafter{\@tempc}}}

\bibitem[\protect\citeauthoryear{{Bickley} et~al.,}{{Bickley} et~al.}{2021}]{Bickley2021}
{Bickley} R.~W.,  et~al., 2021, \mn@doi [\mnras] {10.1093/mnras/stab806}, \href {https://ui.adsabs.harvard.edu/abs/2021MNRAS.504..372B} {504, 372}

\bibitem[\protect\citeauthoryear{{Bickley}, {Ellison}, {Patton}, {Bottrell}, {Gwyn}  \& {Hudson}}{{Bickley} et~al.}{2022}]{Bickley2022}
{Bickley} R.~W.,  {Ellison} S.~L.,  {Patton} D.~R.,  {Bottrell} C.,  {Gwyn} S.,   {Hudson} M.~J.,  2022, \mn@doi [\mnras] {10.1093/mnras/stac1500}, \href {https://ui.adsabs.harvard.edu/abs/2022MNRAS.514.3294B} {514, 3294}

\bibitem[\protect\citeauthoryear{{Bickley}, {Wilkinson}, {Ferreira}, {Ellison}, {Bottrell}  \& {Jyoti}}{{Bickley} et~al.}{2024}]{Bickley2024b}
{Bickley} R.~W.,  {Wilkinson} S.,  {Ferreira} L.,  {Ellison} S.~L.,  {Bottrell} C.,   {Jyoti} D.,  2024, \mn@doi [arXiv e-prints] {10.48550/arXiv.2409.17081}, \href {https://ui.adsabs.harvard.edu/abs/2024arXiv240917081B} {p. arXiv:2409.17081}

\bibitem[\protect\citeauthoryear{{Bottrell} et~al.,}{{Bottrell} et~al.}{2024}]{Bottrell2024}
{Bottrell} C.,  et~al., 2024, \mn@doi [\mnras] {10.1093/mnras/stad2971}, \href {https://ui.adsabs.harvard.edu/abs/2024MNRAS.527.6506B} {527, 6506}

\bibitem[\protect\citeauthoryear{{Casteels} et~al.,}{{Casteels} et~al.}{2014}]{Casteels2014}
{Casteels} K. R.~V.,  et~al., 2014, \mn@doi [\mnras] {10.1093/mnras/stu1799}, \href {https://ui.adsabs.harvard.edu/abs/2014MNRAS.445.1157C} {445, 1157}

\bibitem[\protect\citeauthoryear{{Conselice}, {Bershady}, {Dickinson}  \& {Papovich}}{{Conselice} et~al.}{2003}]{Conselice2003b}
{Conselice} C.~J.,  {Bershady} M.~A.,  {Dickinson} M.,   {Papovich} C.,  2003, \mn@doi [\aj] {10.1086/377318}, \href {https://ui.adsabs.harvard.edu/abs/2003AJ....126.1183C} {126, 1183}

\bibitem[\protect\citeauthoryear{{Duan} et~al.,}{{Duan} et~al.}{2024}]{Duan2024}
{Duan} Q.,  et~al., 2024, \mn@doi [arXiv e-prints] {10.48550/arXiv.2407.09472}, \href {https://ui.adsabs.harvard.edu/abs/2024arXiv240709472D} {p. arXiv:2407.09472}

\bibitem[\protect\citeauthoryear{{Duncan} et~al.,}{{Duncan} et~al.}{2019}]{Duncan2019}
{Duncan} K.,  et~al., 2019, \mn@doi [\apj] {10.3847/1538-4357/ab148a}, \href {https://ui.adsabs.harvard.edu/abs/2019ApJ...876..110D} {876, 110}

\bibitem[\protect\citeauthoryear{{Ellison}, {Patton}, {Simard}  \& {McConnachie}}{{Ellison} et~al.}{2008}]{Ellison2008}
{Ellison} S.~L.,  {Patton} D.~R.,  {Simard} L.,   {McConnachie} A.~W.,  2008, \mn@doi [\aj] {10.1088/0004-6256/135/5/1877}, \href {https://ui.adsabs.harvard.edu/abs/2008AJ....135.1877E} {135, 1877}

\bibitem[\protect\citeauthoryear{{Ellison}, {Mendel}, {Patton}  \& {Scudder}}{{Ellison} et~al.}{2013}]{Ellison2013}
{Ellison} S.~L.,  {Mendel} J.~T.,  {Patton} D.~R.,   {Scudder} J.~M.,  2013, \mn@doi [\mnras] {10.1093/mnras/stt1562}, \href {https://ui.adsabs.harvard.edu/abs/2013MNRAS.435.3627E} {435, 3627}

\bibitem[\protect\citeauthoryear{{Ellison}, {Ferreira}, Wild, Wilkinson, Rowlands  \& Patton}{{Ellison} et~al.}{2024}]{ellison:submitted}
{Ellison} L.~S.,  {Ferreira} L.,  Wild V.,  Wilkinson S.,  Rowlands K.,   Patton R.~P.,  2024, Galaxy Evolution in the Post-merger Regime. II, Submitted

\bibitem[\protect\citeauthoryear{{Ferreira} et~al.,}{{Ferreira} et~al.}{2024}]{Ferreira2024a}
{Ferreira} L.,  et~al., 2024, \mn@doi [arXiv e-prints] {10.48550/arXiv.2407.18396}, \href {https://ui.adsabs.harvard.edu/abs/2024arXiv240718396F} {p. arXiv:2407.18396}

\bibitem[\protect\citeauthoryear{{Garay-Solis}, {Barrera-Ballesteros}, {Colombo}, {S{\'a}nchez}, {Lugo-Aranda}, {Villanueva}, {Wong}  \& {Bolatto}}{{Garay-Solis} et~al.}{2023}]{Garay-Solis2023}
{Garay-Solis} Y.,  {Barrera-Ballesteros} J.~K.,  {Colombo} D.,  {S{\'a}nchez} S.~F.,  {Lugo-Aranda} A.~Z.,  {Villanueva} V.,  {Wong} T.,   {Bolatto} A.~D.,  2023, \mn@doi [\apj] {10.3847/1538-4357/acd781}, \href {https://ui.adsabs.harvard.edu/abs/2023ApJ...952..122G} {952, 122}

\bibitem[\protect\citeauthoryear{{Hani}, {Sparre}, {Ellison}, {Torrey}  \& {Vogelsberger}}{{Hani} et~al.}{2018}]{Hani2016}
{Hani} M.~H.,  {Sparre} M.,  {Ellison} S.~L.,  {Torrey} P.,   {Vogelsberger} M.,  2018, \mn@doi [\mnras] {10.1093/mnras/stx3252}, \href {https://ui.adsabs.harvard.edu/abs/2018MNRAS.475.1160H} {475, 1160}

\bibitem[\protect\citeauthoryear{{Hani}, {Gosain}, {Ellison}, {Patton}  \& {Torrey}}{{Hani} et~al.}{2020}]{Hani2020}
{Hani} M.~H.,  {Gosain} H.,  {Ellison} S.~L.,  {Patton} D.~R.,   {Torrey} P.,  2020, \mn@doi [\mnras] {10.1093/mnras/staa459}, \href {https://ui.adsabs.harvard.edu/abs/2020MNRAS.493.3716H} {493, 3716}

\bibitem[\protect\citeauthoryear{Jr.}{Jr.}{1951}]{KSTEST}
Jr. F. J.~M.,  1951, \mn@doi [Journal of the American Statistical Association] {10.1080/01621459.1951.10500769}, 46, 68

\bibitem[\protect\citeauthoryear{{Kartaltepe} et~al.,}{{Kartaltepe} et~al.}{2015}]{Kartaltepe2015}
{Kartaltepe} J.~S.,  et~al., 2015, \mn@doi [\apjs] {10.1088/0067-0049/221/1/11}, \href {https://ui.adsabs.harvard.edu/abs/2015ApJS..221...11K} {221, 11}

\bibitem[\protect\citeauthoryear{{Kauffmann} et~al.,}{{Kauffmann} et~al.}{2003}]{Kauffmann2003}
{Kauffmann} G.,  et~al., 2003, \mn@doi [\mnras] {10.1111/j.1365-2966.2003.07154.x}, \href {https://ui.adsabs.harvard.edu/abs/2003MNRAS.346.1055K} {346, 1055}

\bibitem[\protect\citeauthoryear{{Koppula} et~al.,}{{Koppula} et~al.}{2021}]{Koppula2021}
{Koppula} S.,  et~al., 2021, \mn@doi [arXiv e-prints] {10.48550/arXiv.2102.05182}, \href {https://ui.adsabs.harvard.edu/abs/2021arXiv210205182K} {p. arXiv:2102.05182}

\bibitem[\protect\citeauthoryear{{Madau} \& {Dickinson}}{{Madau} \& {Dickinson}}{2014}]{Madau2014}
{Madau} P.,  {Dickinson} M.,  2014, \mn@doi [\araa] {10.1146/annurev-astro-081811-125615}, \href {https://ui.adsabs.harvard.edu/abs/2014ARA&A..52..415M} {52, 415}

\bibitem[\protect\citeauthoryear{{Nair} \& {Abraham}}{{Nair} \& {Abraham}}{2010}]{NairandAbraham2010}
{Nair} P.~B.,  {Abraham} R.~G.,  2010, \mn@doi [\apjs] {10.1088/0067-0049/186/2/427}, \href {https://ui.adsabs.harvard.edu/abs/2010ApJS..186..427N} {186, 427}

\bibitem[\protect\citeauthoryear{{Nelson} et~al.,}{{Nelson} et~al.}{2018}]{Nelson2018}
{Nelson} D.,  et~al., 2018, \mn@doi [\mnras] {10.1093/mnras/stx3040}, \href {https://ui.adsabs.harvard.edu/abs/2018MNRAS.475..624N} {475, 624}

\bibitem[\protect\citeauthoryear{{Nikolic}, {Cullen}  \& {Alexander}}{{Nikolic} et~al.}{2004}]{NCA2004}
{Nikolic} B.,  {Cullen} H.,   {Alexander} P.,  2004, \mn@doi [\mnras] {10.1111/j.1365-2966.2004.08366.x}, \href {https://ui.adsabs.harvard.edu/abs/2004MNRAS.355..874N} {355, 874}

\bibitem[\protect\citeauthoryear{{Patton}, {Torrey}, {Ellison}, {Mendel}  \& {Scudder}}{{Patton} et~al.}{2013}]{patton2013}
{Patton} D.~R.,  {Torrey} P.,  {Ellison} S.~L.,  {Mendel} J.~T.,   {Scudder} J.~M.,  2013, \mn@doi [\mnras] {10.1093/mnrasl/slt058}, \href {https://ui.adsabs.harvard.edu/abs/2013MNRAS.433L..59P} {433, L59}

\bibitem[\protect\citeauthoryear{{Patton}, {Qamar}, {Ellison}, {Bluck}, {Simard}, {Mendel}, {Moreno}  \& {Torrey}}{{Patton} et~al.}{2016}]{Patton2016}
{Patton} D.~R.,  {Qamar} F.~D.,  {Ellison} S.~L.,  {Bluck} A. F.~L.,  {Simard} L.,  {Mendel} J.~T.,  {Moreno} J.,   {Torrey} P.,  2016, \mn@doi [\mnras] {10.1093/mnras/stw1494}, \href {https://ui.adsabs.harvard.edu/abs/2016MNRAS.461.2589P} {461, 2589}

\bibitem[\protect\citeauthoryear{{Patton}, {Faria}, {Hani}, {Torrey}, {Ellison}, {Thakur}  \& {Westlake}}{{Patton} et~al.}{2024}]{Patton2024}
{Patton} D.~R.,  {Faria} L.,  {Hani} M.~H.,  {Torrey} P.,  {Ellison} S.~L.,  {Thakur} S.~D.,   {Westlake} R.~I.,  2024, \mn@doi [\mnras] {10.1093/mnras/stae608}, \href {https://ui.adsabs.harvard.edu/abs/2024MNRAS.529.1493P} {529, 1493}

\bibitem[\protect\citeauthoryear{{Pearson}, {Rodriguez-Gomez}, {Kruk}  \& {Margalef-Bentabol}}{{Pearson} et~al.}{2024}]{Pearson2024}
{Pearson} W.~J.,  {Rodriguez-Gomez} V.,  {Kruk} S.,   {Margalef-Bentabol} B.,  2024, \mn@doi [arXiv e-prints] {10.48550/arXiv.2404.11166}, \href {https://ui.adsabs.harvard.edu/abs/2024arXiv240411166P} {p. arXiv:2404.11166}

\bibitem[\protect\citeauthoryear{{Pillepich} et~al.,}{{Pillepich} et~al.}{2018}]{Pillepich2018}
{Pillepich} A.,  et~al., 2018, \mn@doi [\mnras] {10.1093/mnras/stx2656}, \href {https://ui.adsabs.harvard.edu/abs/2018MNRAS.473.4077P} {473, 4077}

\bibitem[\protect\citeauthoryear{{Planck Collaboration} et~al.,}{{Planck Collaboration} et~al.}{2020}]{Planck2018}
{Planck Collaboration} et~al., 2020, \mn@doi [\aap] {10.1051/0004-6361/201833910}, \href {https://ui.adsabs.harvard.edu/abs/2020A&A...641A...6P} {641, A6}

\bibitem[\protect\citeauthoryear{{Privon}, {Barnes}, {Evans}, {Hibbard}, {Yun}, {Mazzarella}, {Armus}  \& {Surace}}{{Privon} et~al.}{2013}]{Privon2013}
{Privon} G.~C.,  {Barnes} J.~E.,  {Evans} A.~S.,  {Hibbard} J.~E.,  {Yun} M.~S.,  {Mazzarella} J.~M.,  {Armus} L.,   {Surace} J.,  2013, \mn@doi [\apj] {10.1088/0004-637X/771/2/120}, \href {https://ui.adsabs.harvard.edu/abs/2013ApJ...771..120P} {771, 120}

\bibitem[\protect\citeauthoryear{{Reeves} \& {Hudson}}{{Reeves} \& {Hudson}}{2024}]{Reeves2024}
{Reeves} A. M.~M.,  {Hudson} M.~J.,  2024, \mn@doi [\mnras] {10.1093/mnras/stad3211}, \href {https://ui.adsabs.harvard.edu/abs/2024MNRAS.527.2037R} {527, 2037}

\bibitem[\protect\citeauthoryear{{Rodriguez-Gomez} et~al.,}{{Rodriguez-Gomez} et~al.}{2015}]{Rodriguez-Gomes2015}
{Rodriguez-Gomez} V.,  et~al., 2015, \mn@doi [\mnras] {10.1093/mnras/stv264}, \href {https://ui.adsabs.harvard.edu/abs/2015MNRAS.449...49R} {449, 49}

\bibitem[\protect\citeauthoryear{{Rodriguez-Gomez} et~al.,}{{Rodriguez-Gomez} et~al.}{2016}]{Rodriguez-Gomez2016}
{Rodriguez-Gomez} V.,  et~al., 2016, \mn@doi [\mnras] {10.1093/mnras/stw456}, \href {https://ui.adsabs.harvard.edu/abs/2016MNRAS.458.2371R} {458, 2371}

\bibitem[\protect\citeauthoryear{{Scott} \& {Kaviraj}}{{Scott} \& {Kaviraj}}{2014}]{Scott2014}
{Scott} C.,  {Kaviraj} S.,  2014, \mn@doi [\mnras] {10.1093/mnras/stt2014}, \href {https://ui.adsabs.harvard.edu/abs/2014MNRAS.437.2137S} {437, 2137}

\bibitem[\protect\citeauthoryear{{Scudder}, {Ellison}, {Torrey}, {Patton}  \& {Mendel}}{{Scudder} et~al.}{2012}]{scudder2012}
{Scudder} J.~M.,  {Ellison} S.~L.,  {Torrey} P.,  {Patton} D.~R.,   {Mendel} J.~T.,  2012, \mn@doi [\mnras] {10.1111/j.1365-2966.2012.21749.x}, \href {https://ui.adsabs.harvard.edu/abs/2012MNRAS.426..549S} {426, 549}

\bibitem[\protect\citeauthoryear{{Snyder}, {Lotz}, {Moody}, {Peth}, {Freeman}, {Ceverino}, {Primack}  \& {Dekel}}{{Snyder} et~al.}{2015}]{Snyder2015}
{Snyder} G.~F.,  {Lotz} J.,  {Moody} C.,  {Peth} M.,  {Freeman} P.,  {Ceverino} D.,  {Primack} J.,   {Dekel} A.,  2015, \mn@doi [\mnras] {10.1093/mnras/stv1231}, \href {https://ui.adsabs.harvard.edu/abs/2015MNRAS.451.4290S} {451, 4290}

\bibitem[\protect\citeauthoryear{{Thorp}, {Ellison}, {Simard}, {S{\'a}nchez}  \& {Antonio}}{{Thorp} et~al.}{2019}]{thorp2019}
{Thorp} M.~D.,  {Ellison} S.~L.,  {Simard} L.,  {S{\'a}nchez} S.~F.,   {Antonio} B.,  2019, \mn@doi [\mnras] {10.1093/mnrasl/sly185}, \href {https://ui.adsabs.harvard.edu/abs/2019MNRAS.482L..55T} {482, L55}

\bibitem[\protect\citeauthoryear{{Violino}, {Ellison}, {Sargent}, {Coppin}, {Scudder}, {Mendel}  \& {Saintonge}}{{Violino} et~al.}{2018}]{Violino2018}
{Violino} G.,  {Ellison} S.~L.,  {Sargent} M.,  {Coppin} K. E.~K.,  {Scudder} J.~M.,  {Mendel} T.~J.,   {Saintonge} A.,  2018, \mn@doi [\mnras] {10.1093/mnras/sty345}, \href {https://ui.adsabs.harvard.edu/abs/2018MNRAS.476.2591V} {476, 2591}

\bibitem[\protect\citeauthoryear{{Wang}, {Pearson}  \& {Rodriguez-Gomez}}{{Wang} et~al.}{2020}]{WPR2020}
{Wang} L.,  {Pearson} W.~J.,   {Rodriguez-Gomez} V.,  2020, \mn@doi [\aap] {10.1051/0004-6361/202038084}, \href {https://ui.adsabs.harvard.edu/abs/2020A&A...644A..87W} {644, A87}

\bibitem[\protect\citeauthoryear{{Wilkinson}, {Ellison}, {Bottrell}, {Bickley}, {Gwyn}, {Cuillandre}  \& {Wild}}{{Wilkinson} et~al.}{2024}]{Wilkinson2024}
{Wilkinson} S.,  {Ellison} S.~L.,  {Bottrell} C.,  {Bickley} R.~W.,  {Gwyn} S.,  {Cuillandre} J.-C.,   {Wild} V.,  2024, \mnras, \href {https://ui.adsabs.harvard.edu/abs/2022MNRAS.516.4354W} {}

\bibitem[\protect\citeauthoryear{{York} et~al.,}{{York} et~al.}{2000}]{York2000}
{York} D.~G.,  et~al., 2000, \mn@doi [\aj] {10.1086/301513}, \href {https://ui.adsabs.harvard.edu/abs/2000AJ....120.1579Y} {120, 1579}

\makeatother
\end{thebibliography}





\bsp	
\label{lastpage}
\end{document}